\documentstyle{mn}
\newcommand{\fma}[1]{\mbox{$#1$}}
\newcommand{\ltsim}{\raisebox{-0.5ex}{$\;\stackrel{<}{\scriptstyle \sim}\;$}}
\newcommand{\gtsim}{\raisebox{-0.5ex}{$\;\stackrel{>}{\scriptstyle \sim}\;$}}
\newcommand{\unit}[1]{\ifmmode \:\mbox{\rm #1}\else \mbox{#1}\fi}
\newcommand{\mone}{\fma{^{-1}}}

\newcommand{\etal}{{et al.\/}}
\newcommand{\eg}{{e.g.\/}}

\newcommand{\cf}{{cf.\/}}
\newcommand{\ha}{H$\alpha$}

\newcommand{\hi}{H~{\sc i}}
\newcommand{\hii}{H\,{\sc ii}}
\newcommand{\hei}{He~{\sc i}}
\newcommand{\heii}{He~{\sc ii}}
\newcommand{\nii}{[N~{\sc ii}]}
\newcommand{\oiii}{[O~{\sc iii}]}
\newcommand{\nai}{Na~{\sc i}}

\newcommand{\kms}{\unit{km~s\mone}}
\newcommand{\cm}{\unit{cm}}

\newcommand{\suns}{\fma{_\odot}}
\newcommand{\msun}{\unit{M\suns}}
\newcommand{\msunyr}{\unit{M\suns~yr\mone}}

\newcommand{\rsun}{\unit{R\suns}}
\newcommand{\flux}{\unit{erg~s\mone\cm$^{-2}$}}

\newcommand{\Mdot}{\fma{\dot M}}
\newcommand{\mdotu}{\fma{\Mdot/u}}
\newcommand{\mdotuten}{\fma{\Mdot/u_{10}}}
\input{psfig}
\begin{document}
\title[Circumstellar H$\alpha$ from Type Ia supernovae]
{Circumstellar H$\alpha$ from SN 1994D and future Type Ia supernovae:
an observational test of progenitor models}

\author[R. J. Cumming et al.]
  {Robert J. Cumming,$^1$
  Peter Lundqvist,$^1$\cr
  Linda J. Smith,$^2$ 
  Max Pettini$^3$ 
  and David L. King$^3$ \\
  $^1$Stockholm Observatory, S-133~36 Saltsj\"obaden, Sweden\\ (e-mail
      robert@astro.su.se, peter@astro.su.se)\\
  $^2$Department of Physics \& Astronomy, University College
      London, London WC1E 6BT, UK\\ (e-mail ljs@star.ucl.ac.uk)\\
  $^3$Royal Greenwich Observatory, Madingley Road, Cambridge 
      CB3~0EZ, UK\\ (e-mail pettini@ast.cam.ac.uk, king@ast.cam.ac.uk)\\
  }
\date{Received 1996 April 3.  Accepted 1996 July.}
\pagerange{\pageref{firstpage}--\pageref{lastpage}}
\pubyear{1996}

\label{firstpage}

\maketitle

\begin{abstract}

Searching for the presence of circumstellar material is currently the
only direct way to discriminate between the different types of
possible progenitor systems for Type Ia supernovae.  We have therefore
looked for narrow \ha\ in a high-resolution spectrum of the normal
Type Ia supernova 1994D taken 10 days before maximum and only 6.5 days
after explosion.  We derive an upper limit of 2.0$\times$10$^{-16}$
\flux\ for an unresolved emission line at the local \hii\ region
velocity.  To estimate the limit this puts on wind density, we have
made time-dependent photoionization calculations.  Assuming spherical
symmetry we find an upper limit of \Mdot\ $\sim 1.5 \times 10^{-5}$
\msunyr\ for a wind speed of 10~\kms.  This limit can exclude only the
highest-mass-loss-rate symbiotic systems as progenitors.  We discuss
the effect of asymmetry and assess the relative merits of early
optical, radio and X-ray limits in constraining mass loss from Type Ia
progenitors.  We find that X-ray observations can probably provide the
most useful limits on the progenitor mass loss, while high-resolution
optical spectroscopy offers our only chance of actually identifying
circumstellar hydrogen.

\end{abstract}


\section{Introduction}

It is generally accepted that a Type Ia supernova (SN Ia) is the
explosion of an accreting white dwarf in a binary system. Surprisingly
though, given how much is known about these events, we still do not
know the nature of the companion star of the progenitor. The issue is
an important one. SNe Ia are visible at cosmological distances and
their luminosities are used to measure the Hubble constant (\eg, Hamuy
\etal\ 1995; Sandage \etal\ 1996) and the deceleration parameter (\eg,
Perlmutter \etal\ 1995).  Our ignorance of Type Ia supernova
progenitors is, furthermore, a gap in our understanding of stellar
evolution in binary systems, and, by virtue of the efficient iron
production of SNe Ia, a stumbling-block to understanding the chemical
evolution of galaxies.

A number of types of system can, in principle, produce a Type Ia supernova.
The most likely (Branch \etal\ 1995) are those where a C-O white dwarf
accretes H or He from a companion, either from its wind or through
Roche lobe overflow, or where two C-O white dwarfs coalesce.  In each
case, the explosion is believed to occur when the progenitor white
dwarf reaches the Chandrasekhar mass, 1.4 \msun.
Sub-Chandrasekhar-mass explosions can also occur if $\sim$0.2 \msun\
of helium is accreted, though current models seem to argue against
their accounting for most SN Ia explosions (Branch \etal\ 1995).

Branch \etal\ (1995) have estimated the likely realisation frequencies
of the different progenitor systems, and conclude that coalescing
pairs of C-O white dwarfs are the most promising candidates.
Symbiotic systems, where the white dwarf accretes H from a red giant
wind, and `hydrogen cataclysmic variables', where H is accreted from a
subgiant through Roche lobe overflow, are also possible.

The latter two cases differ from the first in that hydrogen is
expected to be present in the supernova's circumstellar environment.
A tempting way to discriminate between different types of progenitor
system is therefore to look for signatures of circumstellar matter.
In fact, wherever the white dwarf accretes primarily hydrogen from its
companion, the possibility exists of detecting narrow \hi\ in emission
or absorption, either at early times (Wheeler 1992a,b) or after the
ejecta have become optically thin (Chugai 1986; Livne, Tuchman \&
Wheeler 1992).  If instead helium is accreted, narrow \hei\ or \heii\
lines could be visible (Branch \etal\ 1995).  Radio (Boffi \& Branch
1995) and X-ray (Schlegel \& Petre 1993; Schlegel 1995) observations
might also be able to detect the interaction of the supernova with its
circumstellar medium (CSM).  If the progenitor system is a supersoft
X-ray source, narrow emission lines might be detectable from a
pre-existing ionized nebula (Rappaport \etal\ 1994; Branch \etal\
1995).  Finally, Wang \etal\ (1996) have suggested that polarisation
of the supernova light could reveal the presence of a dusty asymmetric
CSM.

Only a few SNe Ia have shown any evidence at all of a CSM, and in all
cases serious doubt remains.  Branch \etal\ (1983; also Wheeler 1992a)
identified unresolved \ha\ emission in a spectrum of SN 1981B taken 6
days after $B$ maximum (+6 days).  The line is, however, blueshifted
by 2000 \kms\ from the local interstellar Ca~{\sc ii} absorption,
making the identification hard to believe.  Graham \etal\ (1983) and
Graham \& Meikle (1986) reported the detection of an IR excess from SN
1982E.  This supernova was never classified, but exploded in an S0
galaxy, suggesting, at the time, that it was a Type Ia.  Type Ib and
II events have since been seen in S0 galaxies.  Polcaro \& Viotti
(1991) reported narrow \ha\ absorption from SN 1990M at $-$4d, though
an earlier spectrum indicates that the absorption came from the galaxy
rather than the supernova (Della Valle, Benetti \& Panagia 1996).  In
the case of SN 1991bg, no narrow lines were seen at +1d (Leibundgut
\etal\ 1993), but narrow \ha\ and \nai\ were tentatively identified at
+197d by Ruiz-Lapuente \etal\ (1993).  Further spectra, presented by
Turatto \etal\ (1996) show that all the Fe and Co lines characteristic
of SN Ia at this epoch are similarly narrow.  The feature identified
as \nai\ is blended with [Co~{\sc iii}], but there seems to be no
convincing alternative identification for the \ha\ feature (Turatto
\etal\ 1996).

In section \ref{sec-obs}, we report observational limits on early
circumstellar \ha\ from the fairly normal Type Ia SN 1994D.  In
section \ref{sec-discussion} we investigate the implications of \ha\
limits for the progenitor systems of SN 1994D, and Type Ia supernovae
in general.  In particular, we calculate the time-dependent emission
of the \ha\ in terms of a circumstellar interaction model.  Finally,
we assess what can be learnt from early \ha, radio, and X-ray limits.

\section{Observations and results} \label{sec-obs}

On 1994 March 10.14, we used the ISIS spectrograph with an EEV CCD
detector on the William Herschel Telescope to obtain a high-resolution
($R=$35 \kms) spectrum of SN 1994D, centred around \ha.  (The spectrum
was first reported in Cumming, Meikle \& Geballe [1994] and was
briefly presented in King \etal\ [1995].)  The supernova was then only
6.5$\pm$1 days old (H\"oflich 1995), and did not reach $B$ maximum
until 10$\pm$1 days later (Richmond \etal\ 1995; Meikle \etal\ 1996).
The spectrum is thus one of the earliest Type Ia spectra ever taken.
Later high-resolution spectra of SN 1994D have been presented by Ho \&
Filippenko (1995), and by Patat \etal\ (1996).

SN 1994D is located 9.0 arcsec west and 7.8 arcsec north (Richmond
\etal\ 1995) of the nucleus of NGC~4526, an S0$_3$ galaxy in the Virgo
cluster.  We observed with a long slit of width 1.2 arcsec passing
through both the supernova and the galaxy nucleus.  The seeing,
measured from the spatial profile of the supernova, was 1.4 arcsec.
Two integrations of 1000~s and 750~s were added to produce the final
spectrum.  CCD reduction was carried out in the usual way,
wavelength-calibrating with a CuAr arc lamp, and flux-calibrating by
comparison with the flux standard Feige~34 (Stone 1977).  The absolute
flux-calibration is accurate to around 10 percent.

\begin{figure}
\centering
\vspace{6cm}
\includegraphics{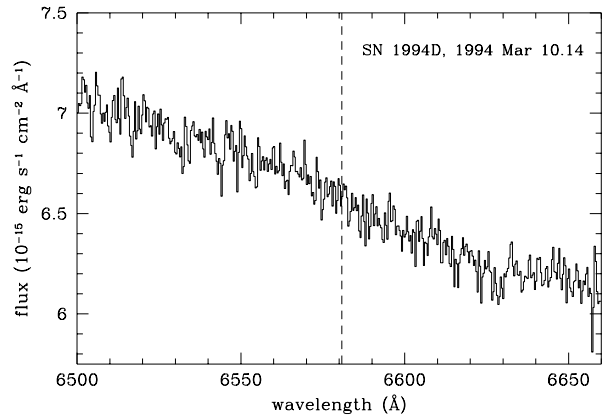}
\caption[]{Spectrum of SN 1994D around \ha.  The broken line marks the
expected velocity of the supernova, +830 \kms.}
\label{f-spectrum}
\end{figure}

The two-dimensional spectra show that faint \hii-region emission in
\ha\ and \nii\ $\lambda\lambda$6548, 6583 can be traced from the
nucleus of NGC~4526 out to the supernova position, but not beyond it.
The line emission shows, however, no abrupt change at the position of
SN 1994D.  At 0.8$\pm$0.3 arcsec from the supernova, the furthest out
from the nucleus we could measure it, the \hii-region \ha\ emission
has a heliocentric velocity $v_{\rm H\,II}=+830\pm50$ \kms.  This is
our best estimate of the radial velocity of the supernova itself.  The
velocities of local interstellar \nai\ absorption, +708 \kms, and the
centre of the galaxy, +625 \kms\ (King \etal\ 1995), are both lower.
Figure \ref{f-spectrum}
shows the high-resolution spectrum around \ha.  The spectrum has not
been dereddened: measurements of the \nai\ interstellar lines by King
\etal\ (1995) and Ho \& Filippenko (1995) give $E_{B-V}$=0.046 and
0.026$^{+0.026}_{-0.013}$, respectively, corresponding to corrections
in flux of only 5$-$10 percent, rather less than other errors in our
results.

The spectrum shows no sign of any narrow \ha\ from the supernova, in
emission or absorption.  (The wavelength range includes no interesting
helium lines.)  We have estimated 3-$\sigma$ detection limits for \ha\
absorption and emission lines of various widths and velocities.  We
did this by adding Gaussian emission and absorption lines of various
strengths, widths and velocities to the observed spectrum, and
measuring the signal-to-noise ratio in the resulting line.

Since detectable circumstellar material is likely to be 
slow-moving, any CSM \ha\ in our spectra would probably be unresolved.
For an unresolved line at the local \hii-region velocity, we found
that the 3-$\sigma$ limit for emission flux is 2.0$\times$10$^{-16}$
\flux.  The corresponding equivalent width limit for an absorption
line is 0.028~\AA.  If we instead allow the true supernova velocity to
differ by up to 200 \kms\ from the local \hii\ region velocity, these
limits increase to 2.7$\times$10$^{-16}$ \flux\ and 0.037~\AA.  The
limits for resolved lines are correspondingly higher.  For $v_{\rm
FWHM}=95$ \kms\ the limits are instead 3.2$\times$10$^{-16}$ \flux\
and 0.062~\AA\ ($v_{\rm SN}$=$v_{\rm H\,II}$), or 5.4$\times$10$^{-16}$
\flux\ and 0.079~\AA\ ($-$200 \kms\ $<(v_{\rm SN}-v_{\rm H\,II})<$ +200
\kms).  The data easily exclude an absorption line of the strength
claimed by Polcaro \& Viotti (1991) from SN 1990M.  Their line had
$v_{\rm FWHM}=760$ \kms\ and an equivalent width of 1.3$\pm$0.2~\AA;
our 3-$\sigma$ limit for a line of this width from SN 1994D is
0.52~\AA.  

\section{Discussion} \label{sec-discussion}

In this section we explore how the \ha\ flux limit for SN 1994D, or
indeed any Type Ia supernova, can be used to constrain models of its
progenitor system.  As circumstellar \ha\ emission requires some
source of excitation, we first assess the different potential sources
of ionizing radiation.  For the most promising of these, we use a
photoionization model to investigate numerically the relationship
between the progenitor system's mass loss rate and the luminosity of
circumstellar \ha.  The model results are used to place a limit on the
mass loss from SN 1994D's progenitor system.  Finally, we compare our
results with early radio and X-ray limits on other supernovae from the
literature, and assess the relative merits of the different techniques
for future events.

\subsection{Ionization sources} \label{sec-i-sources}

We consider four possible ionization sources: ionization by the
radiation accompanying the supernova shock breakout, $\gamma$-rays
from the decay of $^{56}$Ni, the possibility that the CSM is ionized
by the progenitor white dwarf before it explodes (`preionization'),
and radiation from the interaction of the ejecta with a circumstellar
medium.

\subsubsection{Radiation from shock breakout}

The number of ionizing photons produced at shock breakout is probably
quite small. For a black body with a radius of 0.013 \rsun\ at a
temperature of about 10$^6$~K, and the duration of the burst of 1 s
(T. Shigeyama, private communication), we expect only
$\sim$1.5$\times$10$^{48}$ photons with energies above 13.6~eV, enough
only to ionize $\sim$1.3$\times$10$^{-9}$ \msun\ of hydrogen.  We can
therefore exclude this possibility as the source of excitation.

\subsubsection{Radiation due to $^{56}$Ni-decay}

A day or so after the explosion, $\gamma$-rays from $^{56}$Ni-decay in
the supernova ejecta begin to emerge, but these are also unlikely to
ionize the CSM.  Early on, the radiation is too hard, and the UV tail
only shows itself after a few tens of days (P. Pinto, private
communication).  By then, the ejecta will have overrun the densest
parts of the CSM.  We reject this ionzation source as well.

\subsubsection{Preionization} \label{sec-preionization}

To check whether preionization by the progenitor white dwarf is
important, we assume that the CSM is spherically symmetric around the
progenitor, with density $\rho_{\rm wind} \propto r^{-2}$, where $r$
is measured relative to the white dwarf.  This is a reasonable
approximation at radii a few times the separation, $a$, between the
two stars.  For symbiotic systems, $a \sim$10$^{14}$ cm (\eg,
Nussbaumer \& Vogel 1987), and we take this as the minimum radius for
the $r^{-2}$ law.  We can set the density to zero within $r=a$; this
simplification only has a small effect on the results.  We then
compare the rate of ionizing photons from the white dwarf to the total
rate of recombinations in the wind.  For a pure H wind with
temperature $(T_4 \times 10^4)$~K, we find that the the wind is
preionized (at least in the direction away from the red giant) if
\mdotuten\ $\ltsim 5.8\times10^{-8} T_4^{0.48} \dot{N}_{45}^{0.5}
a_{14}^{0.5}$.  Here $u_{10}$ is the wind speed in \kms,
$\dot{N}_{45}$ the rate of ionizing photons emitted by the white dwarf
in units of $10^{45}$~s$^{-1}$, and $a_{14}$ the separation between
the components in units of 10$^{14}$ cm.  Approximating the ionizing
source as a black body of effective temperature $T_{\rm
eff}$=1.5$\times$10$^5$~K with (photospheric) radius $R_{\rm
p}$=9$\times$10$^8$ cm implies that the wind is preionized if
$\mdotuten \ltsim 1.2\times$10$^{-7}T_4^{0.48} a_{14}^{0.5}$.  The
values of $R_{\rm p}$ and $T_{\rm eff}$ are those used for the hot
component of a symbiotic system in the model of Nussbaumer \& Vogel
(1987).  $R_{\rm p}$ is probably somewhat higher than we would expect
for likely Type Ia progenitors, and overestimates the maximum \mdotu\
for which preionization is important.

To estimate the wind temperature in the preionized case we used a
steady-state version of the photoionization code described in
Lundqvist \& Fransson (1988, 1996) and Lundqvist \etal\ (1996, in
preparation).  We adopted $a$=10$^{14}$ cm, $R_{\rm
p}$=9$\times$10$^8$ cm and $T_{\rm eff}$=1.5$\times$10$^5$~K, and
assumed solar abundances for the elements we included (H, He, C, N, O,
Ne, Na, Mg, Al, Si, S, Ar, Ca and Fe).  For
\mdotuten=5$\times$10$^{-8}$, the wind is indeed preionized.  At
$R=10^{15}$ cm, the approximate maximum radius of the ejecta at the
time of our SN 1994D observation, the wind is ionized to, \eg, C~{\sc
v}, N~{\sc v-vi}, O~{\sc v-vi} and the temperature is $\sim$2.5$\times
10^4$~K.  Preionization is only important for $\mdotuten \ltsim
1.9\times$10$^{-7}$, unless $a$ is larger than about $10^{14}$~cm, or
the white dwarf hotter than $T_{\rm eff}$=1.5$\times$10$^5$~K.

\subsubsection{Radiation from the circumstellar interaction region}
\label{sec-cs-radn}

The fourth possibility follows the suggestion of Chevalier (1984) that
Type I SNe whose progenitor white dwarfs have red-giant or AGB-star
companions may interact with their surroundings in a similar manner to
core-collapse supernovae (\eg, SNe 1979C and 1980K: Fransson 1982;
Lundqvist \& Fransson 1988; SN 1993J: Fransson, Lundqvist \& Chevalier
1996).  The supernova ejecta collide with the CSM, setting up a
forward- and reverse-shock structure.  Behind each shock the gas is
heated to high temperatures and produces thermal X-ray emission.  
Being denser, the shocked ejecta radiate more strongly than 
the shocked CSM. 

To check whether the CSM can be ionized by the reverse shock, we have
modelled the circumstellar interaction in a similar way to Chevalier
(1982).  We assume free expansion out to $10^{14}$ cm, and that the
interaction can be described by Chevalier's similarity solutions
beyond that.  The maximum velocity of the ejecta was taken to be
$v_{\rm ej}$=2.0$\times$10$^4$ \kms\ at $10^{14}$ cm, meaning that
this radius is reached in $\sim 0.6$ days.  The density slopes of the
outer ejecta and the wind are described by power-laws, $\rho_{\rm ej}
\propto r^{-n}$ and $\rho_{\rm wind} \propto r^{-s}$, respectively.
We choose $n=7$ (model W7 of Nomoto, Thielemann \& Yokoi [1984]) and
$s=2$.  For solar abundances in the ejecta, the temperature of the
reverse shock $T_{\rm rev} \sim 2.2\times 10^8$~K.  In fact the outer
supernova ejecta are composed of heavy elements like C and O, and the
higher mean molecular weight drives the temperature up to
$\sim$3.5$\times$10$^8$~K.

For \mdotuten=5$\times$10$^{-8}$, and assuming that the shocked ejecta
consist mainly of oxygen (as in W7; Nomoto \etal\ 1984), the ionic
density of the shocked ejecta is $n_{\rm
rev,O}\simeq$1.79$\times$10$^7$ cm$^{-3}$ when the contact
discontinuity between the shocks is at $r=10^{14}$ cm.  The reverse
shock is adiabatic, in contrast to supernovae with larger progenitor
\mdotu.  Behind the reverse shock, the electron-ion equipartition
timescale (\eg, Spitzer 1978) is longer (by a factor of $\sim 4$) than
the time since explosion.  To simulate this we put electron
temperature, $T_{\rm rev,e}$, a factor of 2 lower than the
equipartition temperature.  Assuming the density is the same
throughout the shocked ejecta, and neglecting the frequency-dependence
of the free-free Gaunt factor $g_{\rm ff}$ (we set $g_{\rm ff}$=1.2 at
all frequencies; Rybicki \& Lightman 1979), the outgoing luminosity
from the reverse shock is $L_{\rm rev}(\epsilon) \sim$
5.2$\times$10$^{31}$ $\exp (-\epsilon/kT_{\rm rev,e})$
erg~s$^{-1}$~eV$^{-1}$ at $t_0\simeq0.6$ days.  For $\epsilon\ll kT_{\rm
rev,e}$, $L_{\rm rev}(\epsilon)$ decreases with time as
$(t/t_0)^{-0.6}$, and scales with wind density as (\mdotu)$^2$.


The ability of the radiation from the reverse shock to ionize the wind
depends on the ratio of the ionization timescale, $t_{\rm ion}$, to
the dynamical timescale of the shock, $t_{\rm dyn}=R_{\rm s}/v_{\rm
s}$.  $R_{\rm s}$ and $v_{\rm s}$ are the radius and velocity of
the forward shock.  For our estimated $L_{\rm rev}(\epsilon)$, we find
that $t_{\rm ion}/t_{\rm dyn} \sim 0.50(t/t_0)^{1.2}$ and $t_{\rm ion}
\sim 2.3\times 10^4(t/t_0)^{2.2}$~s for the circumstellar hydrogen
close to the forward shock.  For \mdotuten=5$\times$10$^{-8}$, the
wind does not have enough time to become ionized before being
overtaken by the shock.  At this wind density, therefore, narrow \ha\
emission can only be caused by preionization.

For higher wind densities, the situation is more promising.  Since
$t_{\rm ion}/t_{\rm dyn} \propto (\mdotu)^{-2}$, the wind will be
ionized long enough to allow substantial \ha\ emission.  To
investigate this numerically, we used a time-dependent version of our
computer code with $\mdotuten$ ranging from 5$\times 10^{-8}$ to $1.5
\times 10^{-5}$.  This corresponds to most of the range of plausible
values for $\mdotu$ in symbiotic systems (see section
\ref{sec-limit}).  As before, we assume that the maximum velocity of
the ejecta at $\sim$0.6 days is 2$\times$10$^4$ \kms\ and keep the
approximations described above.  For $\mdotuten \gtsim 2 \times
10^{-7}$, $T_{\rm rev,e}$ actually approaches the equipartition
temperature, so above this value we set $T_{\rm rev,e}=T_{\rm rev}$.
The reverse shock only begins to become radiative at the highest
\mdotuten\ we consider, $1.5 \times 10^{-5}$.  At higher \mdotu\ we
would need to include absorption of the ionizing photons by cool
shocked ejecta (as in SN 1993J; Fransson \etal 1996).

Both the ionized fraction and temperature of the wind increase with
\mdotu.  For example, for \mdotuten=1.5$\times 10^{-7}$ on day 6.5,
the ionized fraction of hydrogen is $\ltsim 0.01$ even close to the
shock, while for \mdotuten=1.5$\times 10^{-5}$ hydrogen is fully
ionized as far out as $\sim$11$R_{\rm s}$.  The temperature close to
the shock in these two cases differs by a factor of $\sim$10
($\sim$2.0$\times 10^4$ and $\sim$2.3$\times 10^5$~K, respectively).
The luminosity-weighted temperature of the \ha-emitting gas varies
less dramatically with \mdotu\ ($\sim$2.0$\times 10^4$ and
$\sim$3.9$\times 10^4$~K), since the \ha\ emission is dominated by
collisional excitation in regions where the temperature is only a few
$\times 10^4$~K.

\begin{figure}
\centering
\vspace{10.5cm}
\includegraphics{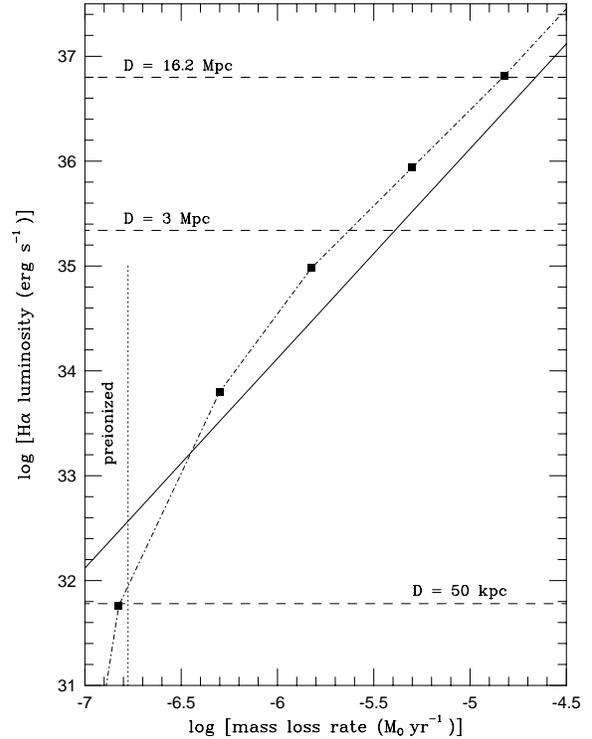}
\caption[]{Luminosity of circumstellar \ha\ emission from a Type Ia
supernova at 6.5 days after explosion, as a function of mass loss from
the companion.  Filled squares show model calculations where the wind
is ionized by the region of circumstellar interaction.  The solid line
shows the corresponding luminosities for a fully-ionized wind at
2$\times$10$^4$~K.  The calculations assume a spherically symmetric
wind with a velocity of 10 \kms, and maximum SN ejecta velocity
$v_{\rm ej}$=2$\times$10$^4$($t$/0.6~d)$^{-1/5}$ \kms.  The vertical dotted line
shows the region where preionization may be important.  The horizontal
dashed lines mark the sensitivity of our SN 1994D observation for a
typical supernova in the Virgo cluster (``16.2 Mpc''), the Local Group
(``3 Mpc'') and the LMC (``50 kpc'').}
\label{f-model}
\end{figure}

The results from our model calculations are shown graphically in 
Figures \ref{f-model}
and \ref{f-lightcurves}.
The solid line in Figure \ref{f-model} shows the \ha\ luminosity at
6.5 days as the result of recombination in a fully-ionized wind at
2$\times$10$^4$~K.  This is close to the temperature which we found for a
preionized wind, and should apply when \mdotu\ is low and the reverse
shock inefficient at ionizing the wind.  The region where
preionization is important is to the left of the vertical dotted line.
The filled squares mark the modelled \ha\ luminosity assuming that the
wind is ionized solely by radiation from the reverse shock.  As
expected, this turns on at around $\mdotuten \sim 2\times 10^{-7}$.
For $\mdotuten \gtsim 4\times 10^{-7}$, the reverse shock excitation
produces more \ha\ than pure recombination does, despite the wind not
being completely ionized.  This is due to efficient collisional
excitation and large optical depths in the Lyman lines.  In
particular, Ly\,$\beta$ is converted into Ly\,$\alpha$ and \ha.
Figure \ref{f-lightcurves} shows how the \ha\ luminosity drops with
time for a range of different \mdotu.

\subsection{Limits on \mdotu\ for SN 1994D and future nearby
supernovae} \label{sec-limit}

We have used the results of our calculations together with our flux
limit on \ha\ from SN 1994D to put a limit on the progenitor system's
\mdotu.  From Figure \ref{f-model} (compare the dashed line marked
`16.2 Mpc' with our numerical results), the limit on an unresolved
emission line at heliocentric velocity 830 \kms\ (section
\ref{sec-obs}) gives $\mdotuten \ltsim 1.5 \times 10^{-5}$ (or $1.6
\times$10$^{-5}$ if $-200 \kms < v_{\rm SN}-v_{\rm H\,II}<200 \kms$).
For a supernova in the local group, at 3 Mpc, the same flux limit
corresponds to $\mdotuten \ltsim 2.5 \times 10^{-6}$ (or
3.3$\times$10$^{-6}$).  As the model predicts a low temperature of the
CSM ($\ltsim 4.0 \times 10^4$~K), we expect the \ha\ line to be
unresolved.

Our limit on \mdotuten\ for SN 1994D's progenitor system only excludes
those symbiotic systems with the very highest mass loss rates.
\mdotuten\ has been measured for about a hundred red giants in
symbiotic systems (\eg, Seaquist \& Taylor 1990; M\"urset \etal\ 1991;
Seaquist, Krogulec \& Taylor 1992).  The values lie in the range
10$^{-8}$ to 2$\times$10$^{-5}$; around two-thirds of the observed
systems have \mdotuten\ within a factor of 4 of 1.0$\times$10$^{-7}$
(Seaquist \etal\ 1992).  Figures \ref{f-model} and \ref{f-lightcurves}
suggest that early \ha\ observations of closer Type Ia events at
comparable sensitivity can detect, or exclude, a fair fraction of
symbiotic systems.  Most, though, seem likely to remain undetectable.
Even a 10\,000-s integration at the WHT on a supernova at 3 Mpc caught
just 3 days after explosion can only detect \ha\ for
\mdotuten$\gtsim10^{-6}$.  

These limits depend, of course, on our assumption of a smooth,
spherically symmetric wind, and on our choice of shock velocity and
power-law index for the ejecta density.   A full investigation of the
parameters is beyond the scope of this paper; we report some further
calculations in Lundqvist \& Cumming (1997).  Our limit on \mdotu\
is probably accurate to within a factor of a few, at least for a
spherical wind.  The effects of asymmetry are harder to assess. 

In principle, our model calculations can also be used to interpret our
limits on \ha\ absorption towards SN 1994D.  The resulting limit on
\mdotu\ would, however, depend more sensitively on the geometry of the
CSM than the emission limit does.

\subsection{Comparison with radio and X-ray limits} 
\label{sec-radio-x}

Limits on \mdotu\ (sometimes quoted as limits on $\Mdot$) have been
obtained for a few SN Ia by placing limits on radio and X-ray
emission.  The radio limit reported by Sramek \& Weiler (1990) for SN
1981B at $\sim$18 days after explosion suggested that \mdotuten\ was
probably less than about 10$^{-6}$ (Boffi \& Branch 1995).  Eck \etal\
(1995) reported that their radio non-detection of SN 1986G (distance
$\sim$4 Mpc) at $-$7 days was in `clear conflict' with the range
$10^{-7} \ltsim \mdotuten \ltsim3\times10^{-6}$.  Observing in X-rays,
Schlegel \& Petre (1993) established a limit for SN 1992A ($\sim$17
Mpc) of (2$-$3)$\times$10$^{-6}$ as late as $\sim$16 days after
\begin{figure}
\centering
\vspace{10.5cm}
\includegraphics{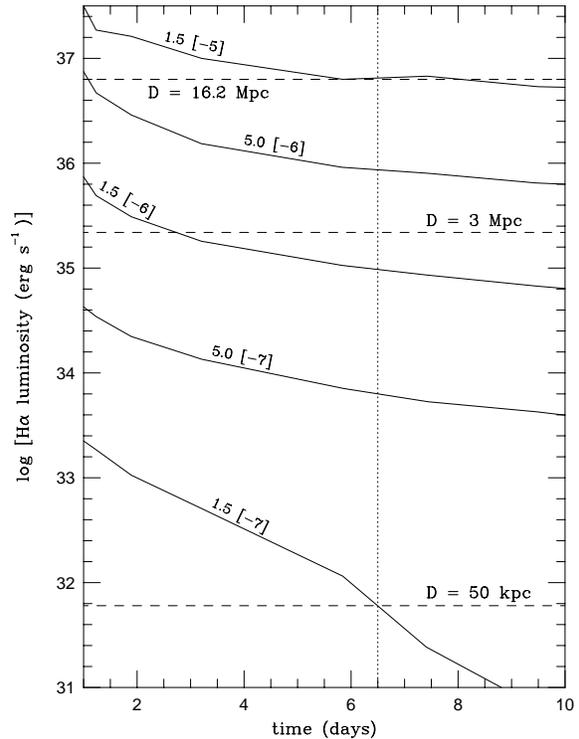}
\caption[]{Evolution of \ha\ luminosity for the model in Figure
\ref{f-model}.  Light curves are shown for \mdotuten\ between
1.5$\times10^{-7}$ (bottom) and 1.5$\times10^{-5}$ (top).  The
vertical dotted line marks 6.5 days after explosion, the epoch of our
SN 1994D observation.  As in Figure \ref{f-model}, the horizontal
dashed lines mark the sensitivity of our SN 1994D observation for
events at different distances.}
\label{f-lightcurves}
\end{figure}
maximum, assuming that any X-rays would arise from circumstellar
interaction.  Limits on \mdotu\ from \ha\ emission scale roughly
linearly with distance.  Our limit for SN 1994D corresponds therefore
to $\mdotuten\ltsim3.3\times10^{-6}$ at the distance of SN 1986G (\cf\
Figure \ref{f-model}).  The radio, X-ray and \ha\ emission all
decrease roughly as $t^{-1}$ after the explosion (see Section
\ref{sec-cs-radn} and Figure
\ref{f-lightcurves}).  This, together with the late epoch of Schlegel
\& Petre's limit suggests that it is X-ray observations which are the
most promising way of detecting the CSM of a Type Ia supernova.  At
all wavebands, observation as soon as possible after discovery is
crucial.

Upper limits on \mdotu\ based, like ours, on the standard Chevalier
model (Chevalier 1982) are probably robust to relaxing the model's
assumption of spherical symmetry.  The presence of its white dwarf
companion means that mass loss from the red giant in a symbiotic
system is unlikely to be spherically symmetric.  Observations of those
few systems which have circumstellar nebulae have generally shown
bipolar structures on both large and small scales (Solf \& Ulrich 1985;
Corradi \& Schwarz 1993; Munari \& Patat 1993).  Aspherical mass loss
necessarily affects the interpretation of limits on radio, X-ray and
optical line emission.  For example, taking optically thin
emission, we estimate that if mass loss were equatorially concentrated
into a solid angle 4$\pi f$, where $f$$<$1, then assuming spherical
symmetry in the Chevalier model will always overestimate the value of
\mdotu\ by a factor of around $f^{-1/2}$, for at least radio and X-ray
observations.  

A problem with radio emission in the Chevalier model is that the radio
flux depends on the efficiency of converting kinetic energy of the
supernova blast wave to magnetic field energy and energy of
relativistic electrons.  This conversion is uncertain by a factor of
at least 10, corresponding to factor of $\gtsim$3 in \mdotu.  If
synchrotron self-absorption is important for radio emission from SN Ia
(as it seems to have been for the early emission from SN 1987A;
Chevalier \& Dwarkadas 1995) then the uncertainty may be even greater.

Limits on early X-ray flux seem therefore likely to give the most
reliable limits on \mdotu\ from the system, since the X-rays probe
the circumstellar interaction directly.  \mdotu\ determined from
radio and optical lines will be uncertain to within a factor of at
least a few.  However, if detected, optical lines could provide much
more powerful constraints on the composition, radiative acceleration,
structure and geometry of the wind than either X-ray or radio
observations.

\subsection{Limits on a supersoft X-ray source progenitor system}
\label{sec-ssxs}

Finally, our spectrum can in principle also test for another possible
class of progenitor, the supersoft X-ray sources (SSXS; \eg, van den
Heuvel \etal\ 1992).  Some SSXS have surrounding photoionized nebulae,
which may still be detectable after explosion (Branch \etal\ 1995),
assuming the recombination timescale is long enough.  We compared our
limits at \ha\ and \nii\ $\lambda$6583 with predictions from Rappaport
\etal\ (1994) and found that the maximum expected line flux in
$\lambda$6583, the stronger of the two lines, is a factor of about
three below our detection limit.  The \ha\ detected from the SSXS
CAL~83 in the LMC (Remillard \etal\ 1995) would have been a factor of
30 below our limit.  Future searches for such emission would be better
carried out around \oiii\ $\lambda$5007, much the strongest optical
line expected from SSXS nebulae.

\section{Summary}

Very early high-resolution optical spectroscopy of SN 1994D shows no
sign of circumstellar \ha.  We have used a numerical photoionization
model to investigate the \ha\ emission assuming that radiation from
circumstellar interaction ionizes the stellar wind from a binary
companion to the progenitor white dwarf.  Comparing our modelling with
the observations we have derived a limit to the progenitor mass loss
which excludes a symbiotic system with a very high mass loss rate.
Future observations of closer Type Ia supernovae can at best detect
systems with mass loss rates around 10$^{-6}$ \msunyr.  Radio, X-ray
and optical searches for circumstellar matter round SN Ia can
complement each other: while X-ray observations can probably set the
most stringent limits on progenitor mass loss, high-resolution optical
spectroscopy offers the exciting possibility of unambiguously
detecting circumstellar hydrogen, if it is present.

\section{Acknowledgements} 

The observational data presented here are available as part of the La
Palma archive of reduced supernova data (Martin \etal\ 1994; Meikle
\etal\ 1995).  We thank Eddie Baron, Francesca Boffi, David Branch,
Claes Fransson, Peter Meikle, Ken Nomoto, Miguel P\'erez-Torres, Phil
Pinto and Toshikazu Shigeyama for helpful discussions during the early
stages of this work, and Jim Lewis for help with accessing the raw
data.  PL acknowledges the financial support of the Swedish Natural
Science Research Council.  LJS acknowledges financial support from
PPARC.

The William Herschel Telescope is operated on the island of La Palma
by the Royal Greenwich Observatory in the Spanish Observatorio del
Roque de los Muchachos of the Instituto de Astrof\'{\i}sica de
Canarias.


\label{lastpage}

\end{document}